\documentclass[fleqn,usenatbib]{mnras}

\usepackage{newtxtext,newtxmath}

\usepackage[T1]{fontenc}
\usepackage{ae,aecompl}

\usepackage{graphicx}	
\usepackage{amsmath}	
\usepackage{amssymb}	
\usepackage[flushleft]{threeparttable}
\usepackage{comment}
\usepackage{multirow}

\newcommand{\ps}{\,s$^{-1}$}
\newcommand{\flux}{\,erg\,s$^{-1}$\,cm$^{-2}$}

\newcommand{\pim}{$\pi$\,Men}

\title[The XUV environment of $\pi$\,Men\,c]{The XUV irradiation and likely atmospheric escape of the super-Earth $\pi$\,Men\,c}

\author[G. W. King et al.]{
George W. King,$^{1,2}$\thanks{E-mail: g.king.5@warwick.ac.uk}
Peter J. Wheatley,$^{1,2}$\thanks{E-mail: p.j.wheatley@warwick.ac.uk}
Vincent Bourrier,$^3$
and David Ehrenreich$^3$
\\
$^{1}$Department of Physics, University of Warwick, Gibbet Hill Road, Coventry, CV4 7AL, UK\\
$^{2}$Centre for Exoplanets and Habitability, University of Warwick, Gibbet Hill Road, Coventry, CV4 7AL, UK\\
$^3$Observatoire de l'Universit\'{e} de Gen\`{e}ve, 51 chemin des Maillettes, 1290 Sauverny, Switzerland
}

\date{Accepted XXX. Received YYY; in original form ZZZ}

\pubyear{2018}

\begin{document}
\label{firstpage}
\pagerange{\pageref{firstpage}--\pageref{lastpage}}
\maketitle

\begin{abstract}
\pim\,c was recently announced as the first confirmed exoplanet from the TESS mission. The planet has a radius of just 2\,R$_{\rm\oplus}$ and it transits a nearby Sun-like star of naked-eye brightness, making it the ideal target for atmospheric characterisation of a super-Earth. Here we analyse archival \textit{ROSAT} and \textit{Swift} observations of \pim\ in order to determine the X-ray and extreme-ultraviolet irradiation of the planetary atmosphere and assess whether atmospheric escape is likely to be on-going. We find that \pim\ has a similar level of X-ray emission to the Sun, with $L_{\rm X}/L_{\rm bol} = (4.84^{+0.92}_{-0.84})\times10^{-7}$. However, due to its small orbital separation, the high-energy irradiation of the super-Earth is around 2000 times stronger than suffered by the Earth. We show that this is sufficient to drive atmospheric escape at a rate greater than that readily detected from the warm Neptune GJ\,436b. Furthermore, we estimate \pim\ to be four times brighter at Ly\,$\alpha$ than GJ\,436. Given the small atmospheric scale heights of super-Earths, together with their potentially cloudy atmospheres, and the consequent difficulty in measuring transmission spectra, we conclude that ultraviolet absorption by material escaping \pim\,c presents the best opportunity currently to determine the atmospheric composition of a super-Earth. 
\end{abstract}




\section{Introduction}
The Transiting Exoplanet Survey Satellite \citep[TESS;][]{Ricker2015} was recently launched by NASA and has begun its search for new exoplanets. In targeting bright, nearby stars, TESS will identify planets that are ideal for follow up observations and further characterisation. Its first confirmed discovery is a small 2\,R$_\oplus$ planet around the $V$=5.7, G0V star \pim\ (HD\,39091) \citep{Huang2018,Gandolfi2018}. The star was already known to host a long-period, non-transiting, eccentric companion with a minimum mass of 10\,M$_{\rm Jup}$ \citep{Jones2002}. As such, the TESS discovery is designated \pim\,c.

\pim\,c will likely prove to be an important discovery, given its relatively small size and bright host. Its size is large enough to suggest the presence of a substantial envelope of volatile material, and the survival of this envelope is consistent with its position above the 
valley in the radius--separation distribution of exoplanets 
that was recently identified observationally \citep{Fulton2017,VanEylen2018}, and suggested to result from the evaporation of volatile envelopes around rocky cores \citep[e.g.][]{Owen2017}. Planets at similarly close separations from their host star have been shown to be undergoing detectable atmospheric mass loss, from Jupiter size \citep[e.g. HD\,209458b and HD\,189733b;][]{VidalMadjar2003,LDE2012} down to Neptune-size \citep[e.g. GJ\,436b;][]{Kulow14,Ehrenreich2015}. The evaporating nature of these planets manifests itself in the form of much deeper transits in the absorption lines of elements being driven off the planet. 

Searches for evaporation signatures from smaller planets have so far proved inconclusive. HD\,97658b is a planet above the evaporation valley, slightly bigger than \pim\,c, and with a slightly wider orbit. Observations suggested a lack of evaporating hydrogen surrounding the planet \citep{Bourrier2017}. 55\,Cnc\,e is another small, nearby planet, although this time below the evaporation valley and thought to be rocky. It also showed a non-detection at Lyman-$\alpha$ \citep{Ehrenreich2012}. Lyman-$\alpha$ observations of Kepler-444 show variations that might be associated with hydrogen escape from its rocky planets \citep{Bourrier2017K444}.

The brightness and proximity of the \pim\ host star, together with the likelihood of \pim\,c retaining a substantial atmosphere, presents a superb new opportunity to search for mass loss from a super-Earth and hence determine the composition of its atmosphere. 

The atmospheric escape of planets is thought to be driven predominately by X-ray and extreme ultraviolet (EUV, together XUV) radiation from the host star
\citep[e.g.][]{Lammer2003,MurrayClay2009,Owen2016}. Thus, determining the XUV environment is a necessary step in assessing the vulnerability of an exoplanet to atmospheric erosion. 

In this letter we present an analysis of archival X-ray observations of the \pim\ system, predominantly observations made with \textit{ROSAT} during the 1990s, together with more recent observations made with \textit{Swift}. We analyse the X-ray spectrum of \pim\ in order to determine the X-ray flux at the location of the super-Earth. We extrapolate this X-ray flux to the EUV band and estimate the mass loss rate from the planetary atmosphere. 
We also search for variations in the X-ray emission of the star across a range of timescales.

\section{Observations}
Seven pointed observations of \pim\ were made with \textit{ROSAT} between 1990 and 1998. We analysed the three observations with more than ten minutes of live exposure time. All three were made with the PSPC detector, and they are outlined in Table~\ref{tab:obs}. The 1991 observations contained four visits spread across one week. \pim\ was also observed with \textit{ROSAT} as part of its All Sky Survey. There are additional serendipitous observations of the system in the \textit{Swift} archive, taken between 2015 December 31 and 2016 January 6, totalling 9.1\,ks of exposure time.

\begin{table}
  \caption{Details of the three \textit{ROSAT} pointed observations with exposure times greater than 10 minutes.}
  \hspace*{-0.7cm}
  \label{tab:obs}
  \begin{threeparttable}
  \centering
  \begin{tabular}{c c c c}
    \hline
    Obs ID & Exp. Time & Start & End \\
           & (s)       & (TDB$^\dagger$) & (TDB$^\dagger$)\\
    \hline
    RP999998A01 & 7061 & 1991-04-18T02:12 & 1991-04-24T04:31\\
    RP999998A03 & 1408 & 1993-04-12T22:38 & 1993-04-12T23:24\\
    RP180278N00 & 856  & 1998-12-12T13:43 & 1998-12-12T13:58\\
    \hline
\end{tabular}
\begin{tablenotes}
\item $^\dagger$ Barycentric Dynamic Time
\end{tablenotes}
\end{threeparttable}
\end{table}

The updated versions of the two discovery papers give parameters that are in broad agreement with each other. We adopted the stellar and planetary parameters from \citet{Huang2018}, and the positional and kinematic information provided by the second \textit{Gaia} data release \citep{GaiaDR2}. 
Furthermore, the parameters should be better constrained as data from more TESS sectors is taken; \pim\ is located close enough to the southern ecliptic pole that it will observed for six months during the primary mission.

\begin{figure}
\centering
 \includegraphics[width=\columnwidth]{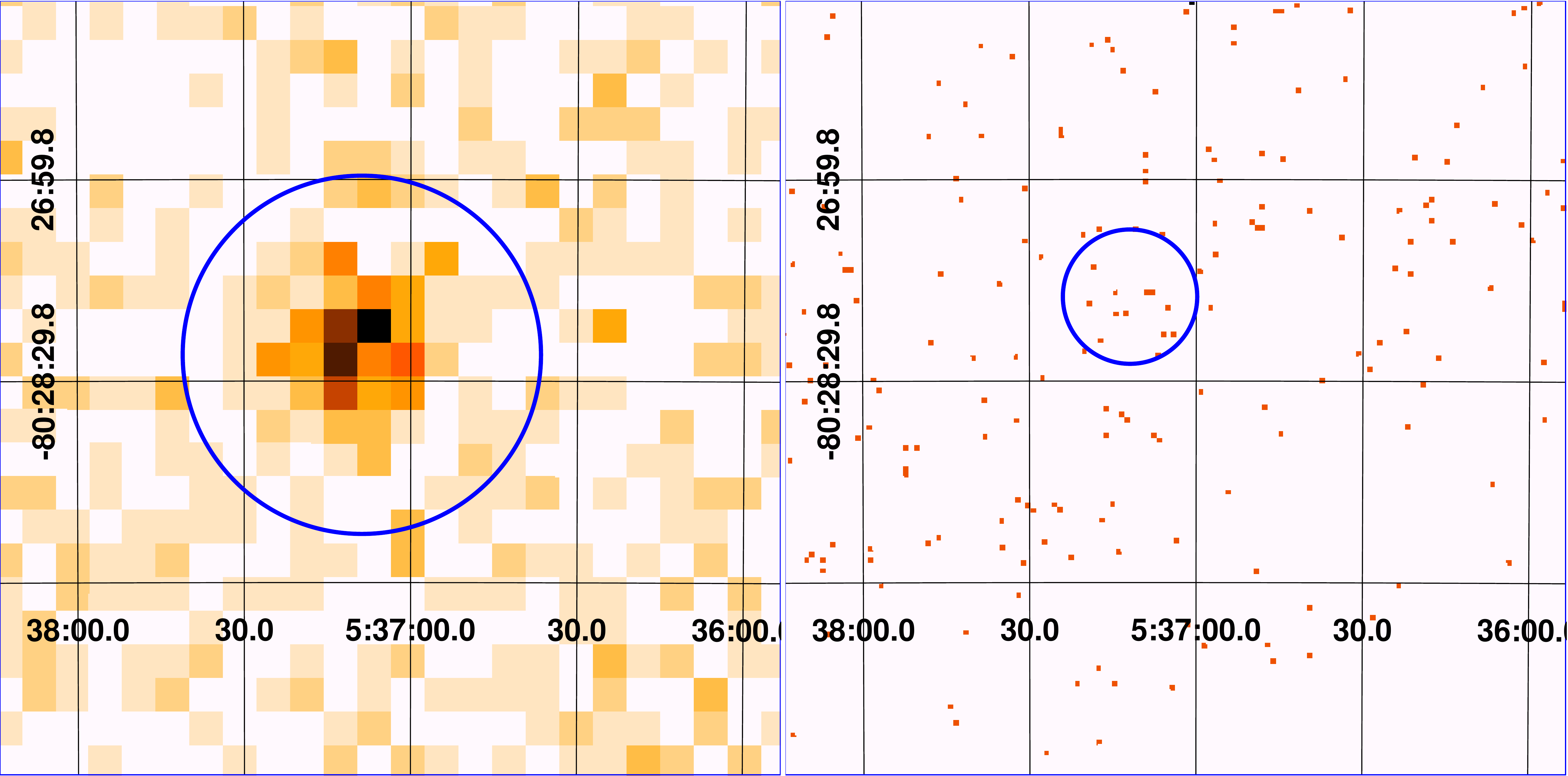}
 \vspace*{-0.5cm}
 \caption{X-ray images of the 1991 \textit{ROSAT} (left), and \textit{Swift} data (right), on the same spatial scale. The blue circles depict the source extraction region used in each case. The single orange points in the right hand plot are individual X-ray photons.}
 \label{fig:xIm}
\end{figure}

All three pointed \textit{ROSAT} observations were performed with the position of \pim\ located on axis. The source was very clearly detected in the longest observations from 1991 (see the image in Fig.\,\ref{fig:xIm}), only marginally detected in 1993, and detected again in 1998. We used an 80\,arcsec radius extraction region for the source, and a single, large 450\,arcsec region for background estimation. Extractions were performed using the \textsc{xselect} program\footnote{https://heasarc.gsfc.nasa.gov/ftools/xselect/}.

For the \textit{Swift} data, we employed 30 and 90\,arcsec source and background regions, respectively. These were again extracted using \textsc{xselect}.

\section{Results}

\subsection{X-ray light curve}

\begin{figure*}
\centering
 \includegraphics[width=\textwidth]{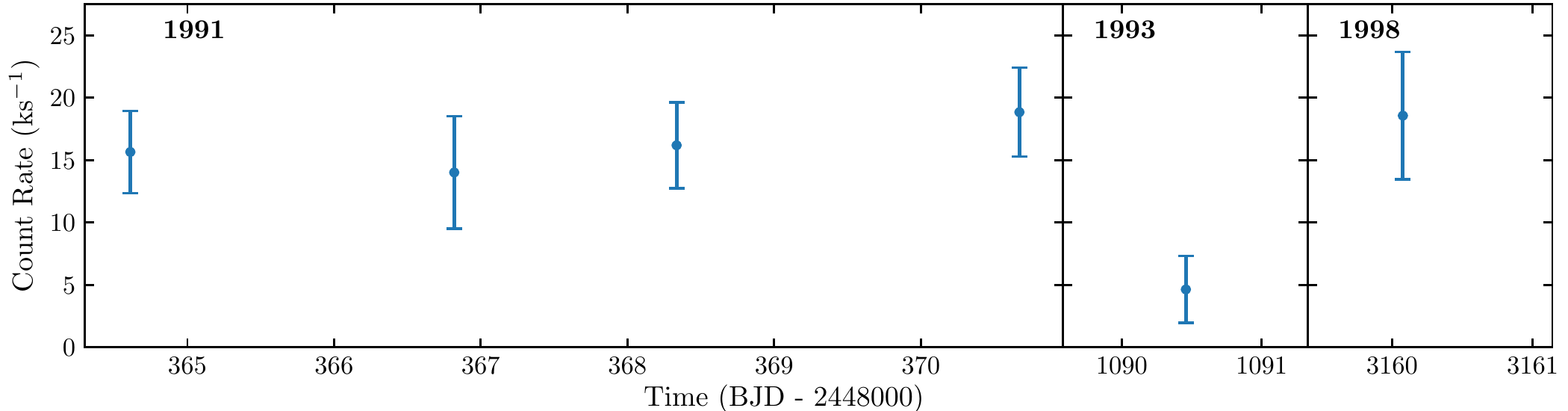}
 \vspace*{-0.6cm}
 \caption{Background corrected \textit{ROSAT} PSPC light curve of \pim\,c, covering the energy range 0.1--2.4\,keV.}
 \label{fig:xlc}
\end{figure*}

The \textit{ROSAT} X-ray light curve of \pim\ is plotted in Fig.~\ref{fig:xlc}, binned to one point per visit. The light curve covers the full PSPC energy range of 0.1 -- 2.4\,keV, although, as discussed in Section~\ref{ssec:xSpec}, the source is very soft with most photons detected having energies below 0.3\,keV.

The single 1993 visit has a count rate considerably below that of the other visits, as suggested by the very marginal detection in that observation. 
The other five visits have count rates that are all consistent with one another to within 1-$\sigma$. Analysis of the All Sky Survey data by us, and that presented in the Second \textit{ROSAT} All Sky Survey source catalogue \citep{Boller2016}, obtain a count rate (22.9$\pm$6.1\,ks$^{-1}$), which is consistent with all of the pointed observations except the 1993 visit. We note that the 1993 visit was not made at the expected time of a planet transit.

\subsection{X-ray spectra}
\label{ssec:xSpec}

\begin{figure}
 \includegraphics[width=\columnwidth]{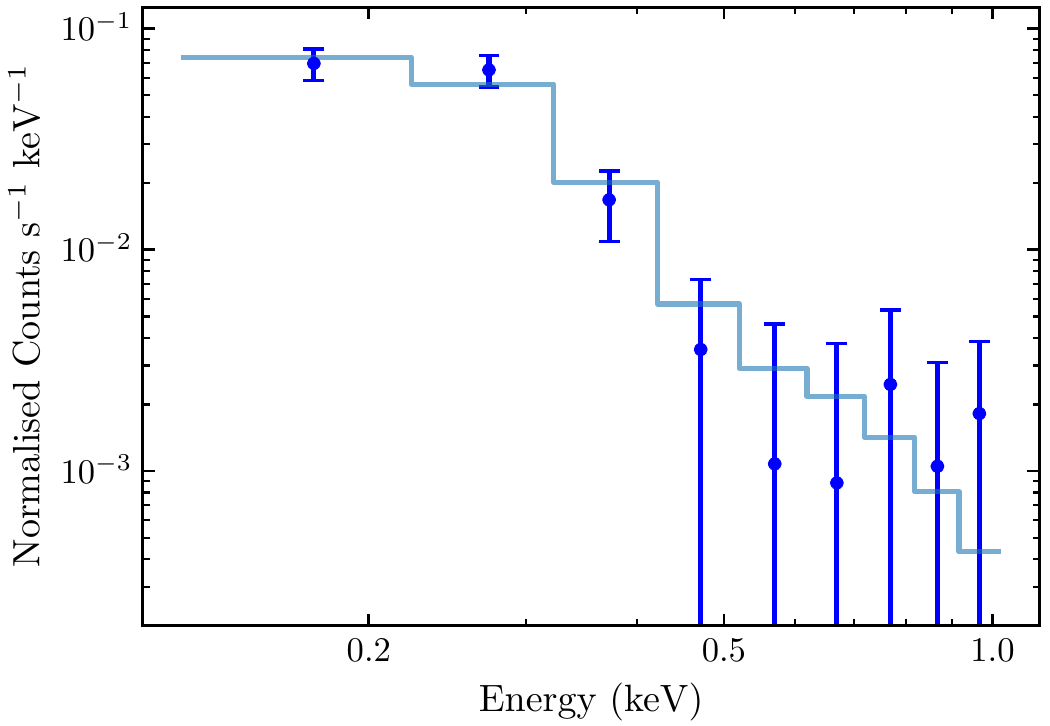}
 \vspace*{-0.6cm}
 \caption{Observed X-ray spectra from the 1991 \textit{ROSAT} observations, displayed along with the best fit model.}
 \label{fig:spec}
\end{figure}

\begin{table}
  \caption{Results of the X-ray spectral analysis, together with fluxes and corresponding planetary irradiation estimates.}
    \hspace*{-0.6cm}
  \label{tab:results}
  	\hspace*{-0.25cm}
  \begin{threeparttable}
  \centering
  \begin{tabular}{l l c}
    \hline\
    Parameter & Symbol & Value \\
    \hline
	Temperature                 & $kT$    		          & $0.152^{+0.029}_{-0.037}$\,keV \\[0.08cm]
	Emission Measure            & EM      		          & $\left( 1.10^{+0.24}_{-0.16} \right) \times 10^{50}$\,cm$^{-3}$ \\[0.08cm]
	Unabsorbed flux at Earth    & $F_{\rm X, \oplus}$     & $\left( 6.7^{+1.3}_{-1.2} \right) \times 10^{-14}$\flux \\[0.08cm]
	X-ray$^a$ luminosity        & $L_{\rm X}$ 			  & $\left( 2.67^{+0.50}_{-0.46} \right) \times 10^{27}$\,erg\ps \\[0.08cm]
	X-ray to bolometric lum.      & $L_{\rm X}/L_{\rm bol}$ & $\left( 4.84^{+0.92}_{-0.84} \right) \times 10^{-7}$ \\[0.08cm]
	EUV$^b$ luminosity          & $L_{\rm EUV}$ 		  & $\left( 21.1^{+4.6}_{-4.3} \right) \times 10^{27}$\,erg\ps \\[0.08cm]
	XUV$^c$ flux at 1\,au       & $F_{\rm XUV, 1\,au}$	  & $8.5^{+1.7}_{-1.6}$\flux \\[0.08cm]
	XUV$^c$ flux at planet c & $F_{\rm XUV, c}$ & $1810^{+350}_{-330}$\,\flux \\[0.08cm]
	XUV received (c.f. Earth)& $P_{\rm XUV, c}$        & $1930^{+390}_{-360}$\,P$_{\rm XUV,\oplus}$ \\
    \hline
\end{tabular}
	\begin{tablenotes}
	\item $^a$ 0.1 -- 2.4\,keV (5.17 -- 124\,\AA); $^b$ 0.0136 -- 0.1\,keV (124 -- 912\,\AA); $^c$ 0.0136 -- 2.4\,keV (5.17 -- 912\,\AA).
    \end{tablenotes}
  \end{threeparttable}
\end{table}

The \textit{ROSAT} PSPC spectrum of \pim\ is displayed in Fig.~\ref{fig:spec}. The spectrum is very soft, and is dominated by photons with energies below 0.3\,keV.

We analysed the spectrum using \textsc{xspec} 12.9.1p~\citep{Arnaud1996}. Our main fit was to the 1991 data only, which has a far larger number of counts compared to the other two epochs. The spectrum was fitted with a single-temperature APEC model, which describes an optically-thin thermal plasma~\citep{Smith2001}. A multiplicative TBABS model was included to account for interstellar absorption~\citep{Wilms2000}. We set the column density of \ion{H}{I} to $5.6\times10^{18}$\,cm$^{-2}$, having assumed a local \ion{H}{I} density of 0.1\,cm$^{-3}$~\citep{Redfield2000}. Abundances were set to Solar values~\citep{Asplund2009}, as [Fe/H] was measured to be close to Solar \citep[$0.08\pm0.03$,][]{Ghezzi2010}. The model was fitted using the C-statistic~\citep{Cash1979} because of the low numbers of counts in some of the higher energy bins.

The best fit model parameters for the 1991 observations, together with their corresponding fluxes and luminosities, are given in Table~\ref{tab:results}. Our analysis uses 0.1 -- 2.4\,keV as the X-ray band and 0.0136 -- 0.1\,keV as the EUV band. The error bars on the fitted parameters and fluxes represent the 68 per cent level, and were estimated using \textsc{xspec}'s MCMC sampler.


Our analysis of the \textit{ROSAT} All Sky Survey data and 1998 visit revealed a similarly soft spectrum, consistent with the 1991 data. 

We extrapolated our X-ray luminosity to the EUV using the empirical relations we determined from Solar \textit{TIMED/SEE} data in \citet{SurveyPaper}, updating that originally presented in \citet{Chadney2015}. Fluxes for the full XUV range, scaled to both the semi-major axis of planet c and 1\,au, are given in Table~\ref{tab:results}. We also present the XUV irradiation scaled to that of the Earth.

\subsection{\textit{Swift} data}
\label{ssec:other}

The XRT instrument on \textit{Swift} extends only down to 0.3\,keV, not low enough to cover the energies where the majority of the \textit{ROSAT} counts were detected. Inspection of the \textit{Swift} data showed only a marginal detection of \pim\ (see the image in Fig.\,\ref{fig:xIm}). We measure a 0.3--2.0\,keV count rate of $0.67\pm0.34$\,ks$^{-1}$ across the 9.1\,ks of exposure time. We estimate there to be 6 source and 3 background counts in the 30\,arcsec aperture. Applying parameters from the 1991 \textit{ROSAT} model fit gives an estimated a \textit{Swift} XRT count rate of 0.76\,ks$^{-1}$, in good agreement with the data.

\section{Discussion}

\subsection{Stellar X-ray emission}

The measured value of $L_{\rm X}/L_{\rm bol}$ for \pim\ of $(4.84^{+0.92}_{-0.84})\times10^{-7}$ is similar to that for the Sun in mid-activity cycle \citep{Judge2003,Ribas2005}. However, the predicted value using the X-ray-age relation of \citet{Jackson2012} is $3\times10^{-6}$, which is almost an order of magnitude greater. Furthermore, using the stellar rotation period of \pim\ \citep[$18.3\pm1.0$\,d;][]{Zurlo2018}, the X-ray-rotation relation of \citet{Wright2011} predicts $5\times10^{-8}$, which is an order of magnitude in the other direction. These discrepancies with empirical age and rotation relations highlight the importance of making direct observations of X-ray irradiation rates of exoplanet host stars. 


The very marginal detection of the star in the 1993 \textit{ROSAT} data is interesting given that the source is detected in the 1998 data with a substantially shorter exposure time (Tab.\,\ref{tab:obs}). 
The X-ray light curve in Fig.\,\ref{fig:xlc} shows that the star is clearly a variable X-ray source. We checked the orbital phase of the planet at the time of the 1993 observations and found that it did not coincide with a planetary transit. The variation is therefore stellar in origin. The consistency of the X-ray brightness across one week in 1991 perhaps indicates that the low flux in 1993 is more likely to be associated with longer timescale variations, perhaps related to a magnetic activity cycle. However, additional X-ray observation are clearly needed in order to determine the variability timescales. 

We estimated the potential count rate for an observation with the EPIC-pn camera on \textit{XMM-Newton}. Applying the best-fit spectral parameters from the fit to the 1991 data yields a count rate of 11\,ks$^{-1}$ with the thick optical blocking filter.


\subsection{Atmospheric escape}
\label{ssec:mLoss}

\begin{table}
  \caption{Estimates of the mass loss rate from the atmosphere of \pim\,c under the different assumptions described in Sect.\,\ref{ssec:mLoss}. $\eta$ refers to the assumed mass loss efficiency and $\beta$ is the effective planet radius at which XUV radiation is absorbed.}
    \hspace*{-1.15cm}
  \label{tab:mLoss}
  \centering
\begin{tabular}{lcccc}
\hline
 Method	& $\eta$	& $\beta$	& $\dot{M}$ & Ref. \\
	& & & ($\times10^{10}$\,g\ps) &	 \\
		\hline
Kubyshkina	& 0.15	& implicit$^a$ & 2.8   & This work	\\
Kubyshkina	& 0.15	& implicit  & 1.2   & \citet{Gandolfi2018} \\
Energy lim.	& 0.15	& 2.67      & 1.5   & This work\\
Energy lim. & 0.15  & 1.00      & 0.11  & This work\\
\hline
\end{tabular}
$^a$ $\beta$ is reported as an output parameter, with a value of 2.67.
\end{table}

The mass loss rate of \pim\,c was previously considered by \citet{Gandolfi2018}, but using an assumed X-ray flux. Those authors employed an interpolation across a set of one-dimensional hydrodynamic models for a hydrogen-dominated atmosphere by \citet{Kubyshkina2018}. We used the same interpolation tool, and used our measured X-ray flux together with the parameters from \citet{Huang2018}. We compare our results with those \citet{Gandolfi2018} in Table~\ref{tab:mLoss}, where it can be seen that  our estimate using the measured XUV flux is a factor of 2.3 higher. 
As discussed by \citet{Gandolfi2018}, such a high mass loss rate implies that \pim\,c has either lost its hydrogen envelope, and now has an atmosphere dominated by heavier elements, or it must have formed with a thick hydrogen atmosphere that has only partly survived.

We also estimated the mass loss rate of \pim\,c, applying the energy-limited method \citep{Watson1981,Erkaev2007}, given by
\begin{equation}
\dot{M} = \frac{ \beta^2 \eta \pi F_{\text{XUV}} R^3_{\text{p}} }{ G K M_{\text{p}} },
\label{eq:massloss}
\end{equation}
where $R_{\rm p}$ and $M_{\rm p}$ are the radius and mass of the planet, respectively. $K$ is a Roche-lobe correction~\citep{Erkaev2007}. This method has been previously applied by numerous studies~\citep[e.g.][]{Salz2015,Louden2017,SurveyPaper}. The evaporation efficiency, $\eta$, was assumed to be a canonical value of 0.15, as also assumed in the \citet{Kubyshkina2018} model. For a discussion of this choice see \citet{SurveyPaper}. 
For the effective planet radius at which XUV radiation is absorbed, $\beta$, we made two choices. First, the value of 2.67 from the hydrodynamical calculation of \citet{Kubyshkina2018}, which is appropriate for a hydrogen-dominated atmosphere. Second, a value of 1.0, corresponding to the limiting case for an atmosphere consisting of heavier species (e.g.\ water or methane). The energy-limited mass loss estimates are also
given in Table~\ref{tab:mLoss}, and we note that for the same $\beta$ the energy limited rate is within a factor 2 of the mass loss rate derived from the hydrodynamical model.

\subsection{Detecting evaporation}

Our predicted mass loss rates for \pim\,c, given in Table~\ref{tab:mLoss} are substantial, with significant implications for the evolution of the planetary atmosphere.   
These escape rates are even higher than our XUV estimates for the Neptune-sized planet GJ\,436b \citep{SurveyPaper}, for which ultraviolet absorption from the escaping atmosphere has been detected \citep{Kulow14,Ehrenreich2015}.
In that case, Ly\,$\alpha$ absorption has been observed up to 56 per cent deep during transits that last for up to 20 hours after the optical transit \citep{Ehrenreich2015,Lavie2017}. 
This deep absorption was found to be consistent with neutral hydrogen escape of only $\left(2.5\pm1\right) \times 10^8$\,g\ps\ \citep{Bourrier2015,Bourrier2016}. 

This favourable comparison with GJ\,436b, together with a bulk density requiring a volatile envelope, suggests that atmospheric escape from \pim\,c should be readily detectable using the \textit{Hubble Space Telescope}: either at Ly\,$\alpha$ or wavelengths associated with heavier species. 
Our predicted atmospheric escape for \pim\,c is 
also 
greater than our estimate for HD\,97658b \citep{SurveyPaper}
for which Ly\,$\alpha$ absorption was not detected \citep{Bourrier2017}.

The species detected/not detected in an extended or escaping atmosphere around the super-Earth would determine the composition of the planetary atmosphere. For example, the presence of both hydrogen and oxygen would point to a H$_2$O rich world, which is consistent with its density from \citet{Huang2018}. Alternatively, hydrogen and helium detections would suggest a substantial gaseous envelope around a dense rocky core, also consistent with the density from \citet{Huang2018}.

The proximity of \pim\ to Earth means that the star should be bright enough in Ly\,$\alpha$, and the interstellar absorption low enough, for a sensitive search for escaping neutral hydrogen. Using the empirical relations of \citet{Linsky2014} linking Ly\,$\alpha$ and EUV fluxes, we use our EUV flux to estimate the Ly\,$\alpha$ flux at Earth of \pim\ to be $8.7\times 10^{-13}$\flux. This is four times that of GJ\,436 \citep{Bourrier2016,Youngblood2016}, and twice that of HD\,97658 \citep{Youngblood2016,Bourrier2017}.

Sensitive searches for other elements and ion species surrounding \pim\,c can also be made. 
Notably, the 10830\,\AA\ helium line that was recently detected for WASP-107b \citep{Spake2018}, as well as ultraviolet lines of carbon, oxygen, nitrogen and magnesium previously detected around hot Jupiters
\citep[e.g.][]{VidalMadjar2004,Fossati2010,Linsky2010,BenJaffel2013}.

\section{Conclusions}
The first confirmed exoplanet from the \textit{TESS} mission, \pim\,c, is a super-Earth orbiting a 
Sun-like star of naked-eye brightness. We find that the star has a soft X-ray spectrum and an X-ray luminosity similar to that of the Sun. It is also a variable X-ray source. We show that \pim\,c suffers XUV irradiation around 2000 times stronger than that of the Earth. 
As a consequence, the planet atmosphere is likely to be escaping at a rate greater than that readily observed for the warm Neptune GJ\,436b. Furthermore, we predict that \pim\ is four times brighter than GJ\,436 at Ly\,$\alpha$. We conclude that the detection of material escaping \pim\,c using ultraviolet and infrared spectroscopy presents our current best opportunity to determine the composition of a super-Earth atmosphere.

\section*{Acknowledgements}
G.W.K. is supported by an STFC studentship (Award number: 1622607). P.J.W. is supported by an STFC consolidated grant (ST/P000495/1).




\bibliographystyle{mnras}
\bibliography{piMen} 








\bsp	
\label{lastpage}
\end{document}